\documentstyle[12pt,epsfig]{article}
\title{Confidence belts on bounded parameters}  
\author{J.Bouchez
\\DAPNIA/SPP
\\CEA-Saclay\\ 91191 Gif-sur-Yvette Cedex, France}
\begin{document}           

\maketitle                 
\begin{abstract}
We show that the unified method recently proposed by Feldman and
Cousins to put confidence intervals on bounded parameters cannot avoid
the possibility of getting null results. A modified bayesian approach is also
proposed (although not advocated) which ensures no null results 
and proper coverage.
\end{abstract}
\section{Introduction}
In a recent paper \cite{CF} , Feldman and Cousins have revisited the
long-standing problem of confidence belts on bounded parameters, for 
which the standard method proposed by Neyman \cite{Ne} leads in some
cases to null, or unphysical, results, and have proposed a new
method.

The advocated method takes advantage of the freedom left by the
Neyman construction in the choice of ordering used to select the
ensemble of values of the measurement with a given probability content.

This new method is strictly classical -or frequentist- (that is, not
using bayesian extensions to classical statistics), avoids overcoverage,
gives a natural and non-biasing transition between upper limits and
intervals, and, according to the authors, avoids null results.

We show in this paper that the last point is not strictly met and that
null results can only be avoided for confidence levels above a limit
which depends upon the probability law under consideration.

\section{The "problem" and its solutions}
\subsection{Building confidence domains}
Let us consider a random variable $x$ of density probability
$f(x|\mu)$ where $\mu$ is an unknown parameter. Given an observation $x$,
one wishes to make a statement on $\mu$ with a given confidence level
(noted CL in the following) $\alpha$. $\alpha$ is the probability that
the statement is true.

Let us first consider the case of an upper limit on $\mu$. The
Neyman construction consists in defining for each value of $\mu$
the value $x_m$ such that
\begin{equation}
  F(x_m|\mu) = \int_{-\infty}^{x_m} f(x|\mu)\:dx = 1 - \alpha
\label{eq:1}
\end{equation}
Whatever $\mu$, $x$ has a probability $\alpha$ to be bigger than $x_m(\mu)$.
$x$ being observed,
the $\alpha$ CL limit $\mu_M$ on $\mu$ is obtained by solving 
$x_m(\mu_M) = x$. In the following, we will consider only the cases where
this equation has a unique solution, implying that $x_m(\mu)$ is a
monotonic increasing function of $\mu$ for any value $\alpha$.
This is equivalent to stating that
\begin{equation}
 \frac{\partial F(x|\mu)}{\partial \mu} < 0\; \vee x,\mu
\label{eq:2}
\end{equation}
One can equally define an $\alpha$ CL interval $[\mu_m,\mu_M]$ on $\mu$
by constructing for each $\mu$ an interval $[x_m(\mu),x_M(\mu)]$ in $x$ of
probability content $\alpha$ and, given the observation $x$,
solve $x_m(\mu_M) = x_M(\mu_m) = x$. But contrary to upper (or lower)
limits, the choice of interval in $x$ is not unique and an ordering
prescription on $x$ is necessary.

The usual prescription consists in defining $x_m$ and $x_M$ by
\begin{equation}
  \int_{-\infty}^{x_m} f(x|\mu)\:dx = \int_{x_M}^{\infty} f(x|\mu)\:dx =
(1 - \alpha)/2
\label{eq:3}
\end{equation}
(the so-called central interval).
This only works for 1-dimensional $x$, and a new ordering for n-dimensional
$x$ has to be chosen. One generally uses the $\chi^2$ between $x$ and
the mean value of $x$, $\bar{x}(\mu)$, or the likelihood ratio
$R = f(x|\mu)/f(x_0|\mu)$ where $f(x_0|\mu)$ is the maximum of $f$, and one
defines a cut $c$ on $\chi^2$ (resp R) so that the probability of
$\chi^2 < c$ (resp $R < c$) is $\alpha$. These two methods (among others), 
when applied to the 1-dimensional case, would generally give non central
intervals.
\subsection{The case of bounded parameters}
The method outlined above can lead to null results on the parameter $\mu$
if this parameter is bounded ( for example, $\mu > 0$ if $\mu$ is a mass,
a variance, etc...). Such a case is illustrated on figure~1, where one sees
that for low enough values of $x$, no upper limit on $\mu$ is obtained (or a
negative limit is obtained in case $f(x|\mu)$ is defined even for negative
$\mu$'s).
\begin{figure}
\begin{center}
\mbox{\epsfig{file=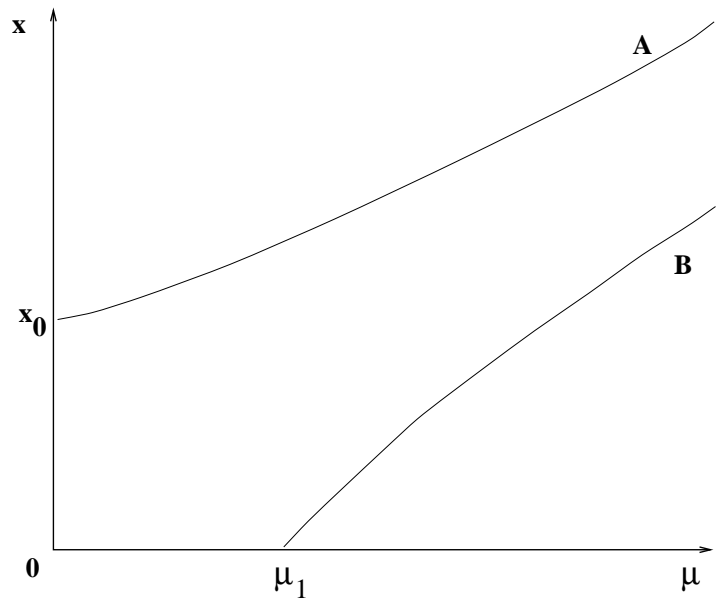,height=8cm}}
\end{center}
\caption{As a fuction of the parameter $\mu$, we show the acceptance domain
in $x$ obtained by the frequentist approach (above curve A defined 
by equation~1) and by the bayesian approach (above curve B defined by
equation~5). With curve A, null results are obtained for $x < x_0$, while
curve B gives overcoverage.}
\end{figure}

For a statistician, this is not a problem, since an $\alpha$ CL statement
has a $(1-\alpha)$ probability to be wrong. For unbounded parameters, it is
impossible to know if a statement is true or false. For bounded parameters,
a fraction of the physicists emitting a wrong statement are aware that
their statement is wrong, and would rather like not to be in this
uncomfortable situation. One should stress however that their result,
which when expressed as a limit on $\mu$ seems to bring no information, is
as legitimate and useful as any other result obtained by those physicists
publishing a physical limit (which might be right or wrong).

But the discomfort caused by such a situation is so strong that cures have
been searched for to avoid publishing null results.

\subsection{The bayesian solution}

The bayesian extension to classical statistics consists in building 
a probability density on the unknown parameter $\mu$ from the
observation $x$ by applying the (classical) Bayes theorem on conditional
probabilities, af if $\mu$ were a random variable \cite{Bayes}.
This gives a  density probability (a degree of belief in bayesian language)
\begin{equation}
 g(\mu|x) = K(x)\:f(x|\mu)\:P(\mu) 
\label{eq:4}
\end{equation}
where K is a normalization coefficient insuring that
\[\int_{-\infty}^{\infty} g(\mu|x)\:dx = 1\]
and $P(\mu)$ summarizes our knowledge
on $\mu$ prior to the observation $x$. In particular, for bounded parameters,
$P(\mu)$ will be null outside the physical region.

An upper limit at "$\alpha$ CL" $\mu_0$ can then be built on $\mu$ by solving
\begin{equation}
\int_{-\infty}^{\mu_0} g(\mu|x)\:dx = \alpha
\label{eq:5}
\end{equation}

By construction, the selected range of physical $\mu$ values will never be
empty, so that
null results are impossible. This is shown on figure~1.

This approach, although recommended by PDG \cite{oldPDG} till 1997, has 
been criticized by
frequentist statisticians for the following reasons:
\begin{enumerate}
\item Near the parameter boundary, where null results happen with the standard
method, this bayesian "$\alpha$ CL" limit 
has a classical CL higher than
$\alpha$ (it is actually 1 if $\mu$ happens to be below $\mu_1$, 
see figure~1), leading to overcoverage and loss of predictive power.
Note however that this overcoverage is a local property, 
and could well
transform into undercoverage
at higher $\mu$ values.
\item There is some arbitrariness in the choice of $P(\mu)$. PDG
proposes to use a Heavyside function to describe the physical limit
on $\mu$. Such a prescription has some drawbacks: it is not invariant
under changes of parametrization, so that limits will actually depend
on the parametrization used. Furthermore, there are cases where
the integral in (\ref{eq:4}) 
is divergent unless $P(\mu)$ is adequately chosen.
\end{enumerate}
I would like , in the next section, to propose a cure to the second point.
\subsection{A modified bayesian limit}
A classical upper limit on $\mu$ set at $\mu_0$ from an observation $x$
has a confidence level $\alpha$ equal to $1-F(x|\mu_0)$. If one is
willing to interpret $P(\mu < \mu_0) = \alpha$ as a probability statement 
on $\mu$, then the cumulative probability distribution for $\mu$ has to
be $1-F(x|\mu)$, so that the probability density for $\mu$ deduced from
the observation $x$ is given by:
\begin{equation}
 \hat{g}(\mu|x)= -\frac{\partial}{\partial \mu}F(x|\mu)
\label{eq:6}
\end{equation}
Contrary to the usual bayesian definition, $\hat{g}$ is always defined
and normalized to 1 by construction. 
Furthermore, when equation \ref{eq:2} is satisfied, which we have supposed,
$\:\hat{g}$ is positive.
When $\mu$ is bounded, the definition of conditional probability 
can be applied to $\hat{g}$ to get its restriction to
physical values. If $\mu > a$, then (dropping $x$ in the notation):
\begin{equation}
 \tilde{g}(\mu|\mu>a) = 
\frac{\hat{g}(\mu)}{\int_{a}^{\infty} \hat{g}(\mu)\:d\mu}
\label{eq:7}
\end{equation}
will be the probability distribution of $\mu$ restricted to its physical
values.
$\tilde{g}$ can then be used instead of g (defined in equation~\ref{eq:4})
to set an upper limit $\mu_0$ on $\mu$, by solving:
\begin{equation}
\int_{a}^{\mu_0} \tilde{g}(\mu)\:d\mu = \alpha
\label{eq:8}
\end{equation}

Using $\tilde{g}$ rather than $g$ has several advantages:
\begin{itemize}
\item When $\mu$ is not bounded, the obtained limit is identical to the
classical limit (for bounded parameters, it can be proven it gives 
overcoverage whatever
the value of $\mu$ is, contrary to the usual bayesian method).
\item the limit is invariant under reparametrization. If we change 
$\mu$ to $\lambda = r(\mu)$, the limit $\lambda_0$ will be $r(\mu_0)$
\item When prior knowledge is limited to the physical boundary, there
is no need to invoke the ambiguous function $P(\mu)$, which is replaced
by a Heavyside function irrespective of the parametrization.
\end{itemize}

Let us note before closing this section that $\hat{g}(\mu|x)=f(x|\mu)$ in 
some special cases. Among them are the normal law of mean $\mu$ and constant
variance, and the Poisson law of mean $\mu$, so that for these cases,
the limit given by $\tilde{g}$ becomes identical to the former
PDG recommendation.
\subsection{Feldman and Cousins solution}
Feldman and Cousins wanted to fulfill 3 conditions:
\begin{enumerate}
\item Avoid null results
\item keep a frequentist approach and produce results with a classical
CL equal to $\alpha$ with no overcoverage (except when $x$ is discrete,
since this discreteness implies unavoidedly some overcoverage)
\item Solve the "flip-flop" problem of how to switch from upper limit
to confidence interval, as they have shown that a choice made according
to the value of the observation $x$ leads, as is usually the case,
to a biased result, namely an undercoverage (the actual CL being lower
than the claimed one). This problem occurs in practice for parameters bounded
from below, $\mu \geq a $, where $\mu -a$ is the strength of an hypothetic
signal on which to make a statement, and observations also bounded from below
($x > x_0$).
\end{enumerate}
These authors propose an ordering principle on $x$ based on
\begin{equation}
r(x) =  \frac{f(x|\mu)}{f(x|\mu_{best})} 
\label{eq:9}
\end{equation}
where $\mu_{best}$ is the physical value of $\mu$ which maximizes 
the denominator. $r(x)$ varies between 0 and 1, and one selects for each
$\mu$ the values of $x$ such that $r(x) > r_c$  and 
\[  \int_{r>r_c} f(x|\mu) dx = \alpha \]
This construction, being classical, meets the second condition.
It will meet the first and third conditions
{\em in cases where} the algorithm selects
$x$ values between $x_{min}(\mu)$  and $x_{max}(\mu)$ such that
$x_{min} = x_0$ for $\mu < a + s_0$, and $x_{min} > x_0$ for $\mu > a + s_0$,
so that an upper limit is published when the observed value of $x$ is below
$x_{max}(a)$, and an interval on $\mu$ (not necessarily central)
for higher values of $x$.

If it is not the case, the algorithm will lead to null results:
\begin{itemize} 
\item If $x_0$ is selected for all values of $\mu$, the 
algorithm will produce lower limits for $x > x_{max}(a)$ and "null" results
below (all physical values of $\mu$ being accepted).
\item If $x_0$ is excluded for all values of $\mu$, it will give a null
result for \newline
$x < x_{min}(a)$, an interval for $x > x_{max}(a)$ and an
upper limit inbetween.
\end{itemize}
We show in the next section that such a situation is likely to occur.
\subsection{Study of the ordering algorithm}
In order to prove the above mentionned fact, it is sufficient to
exhibit cases where null results can be obtained.

I will restrict in the following to cases where $\mu$ and $x$ both take
only positive values, and where $f(x|\mu)$ is , up to a normalization
factor, a function $g(y)$ with $y=x/\mu$. If $\int g(y) dy = 1$, the
normalization factor is $1/\mu$, so that $f(x|\mu) = g(y)/\mu$.
(Such cases can be met in real cases: $y$ can follow a $\chi^2$ law and
$\mu$ be the unknown variance, $x$ then being the unnormalized $\chi^2$
from which one wishes to give a statement on the variance).
Note that the limits on $x$ at any confidence level are just straight
lines passing thru the origin, and no problem of null results is
encountered with the usual classical method. The problems appear only when
for some reason, one restricts $\mu$ to values higher than $a$, a strictly
positive constant (in the $\chi^2$ example, one knows that the variance
is at least $a$, and one wishes to know if there is some other 
contribution to it). Note also that such distributions satisfy the
condition of equation~\ref{eq:2}.

When no bound is imposed on $\mu$ (other than being positive), $\mu_{best}$
as defined in (\ref{eq:9}) is $\mu_0$ given by 
\[\frac{\partial }{\partial \mu} f(x|\mu_0) = 0\]
This equation is equivalent to
\[\frac{d}{dy}h(y_0) = 0\]
where $h(y) = y g(y)$ and $y_0 = x/\mu_0$.

Let us suppose that $h(y)$ is such that it has a unique maximum at $y=y_0$.

To build $r(x)$ when $\mu > a$, two cases have to be considered:
\begin{itemize}
\item when $x > a y_0$, $ \mu_0 > a$, $\mu_{best} = \mu_0$ 
and $r(x) = h(y)/h(y_0)$
\item when $x < a y_0$, $ \mu_0 < a$, $\mu_{best} = a$ 
and $r(x) = h(y)/h(\mu y/a)$
\end{itemize}

Note that $r(x_0=\mu y_0) = 1$, and 
\[r(0) = \lim_{y \rightarrow 0} \frac{h(y)}{h(\mu y/a)} = \left( \frac{a}{\mu}
\right)^{(n+1)}\] where $n$ is the first non-null derivative of 
$g(y)$ at $y=0$.

When $\mu \rightarrow a$, $r(x) \rightarrow 1$ for $x$ between 0 and $ay_0$,
then decreases.

$r(x)$ is shown on figure~2. We will suppose in the following that $r(x)$ is
a non decreasing function of $x$ for $x$ 
between 0 and $a y_0$, as it is the case
in the explicit examples below (this is to avoid disconnected acceptance
domains).
\begin{figure}
\begin{center}
\mbox{\epsfig{file=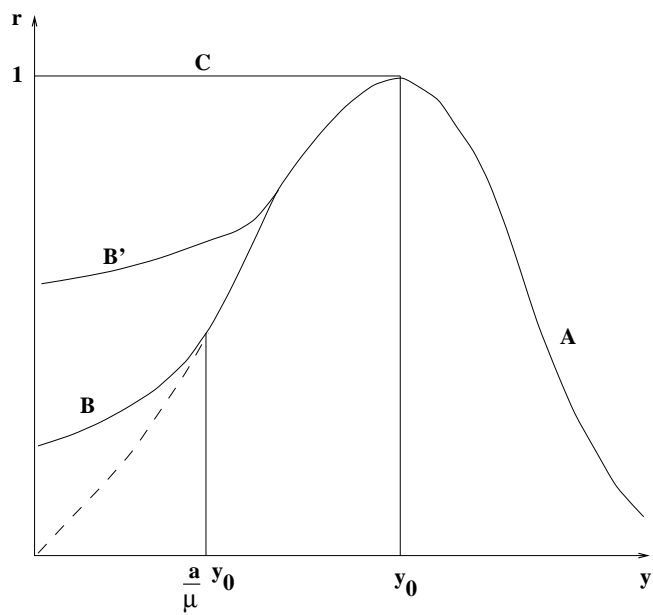,height=8cm}}
\end{center}
\caption{$r$ is shown as a function of $y$. $h(y)$ takes its maximum at
$y_0$. The function $r$ is made of 2 curves, curve A given by $h(y)/h(y_0)$
when $x > a y_0$, curve B given by $h(y)/h(\mu y /a)$ when $x < a y_0$.
The curve B' corresponds to a value $\mu'$ of $\mu$ closer to $a$, and
curve C is obtained when $\mu = a$.}
\end{figure}

Then, the acceptance domain for $x$ contains $ x_0 =\mu y_0$ and 
extends to both sides
according to the chosen value of $\alpha$, including or not $x = 0$.
When $\mu \rightarrow a$, the acceptance domain will or not contain $x=0$
depending on the value of $I = \int_0^{y_0} g(y) dy$.
$I$ is $\leq 1$.
If $I$ is smaller than $\alpha$, the acceptance domain at $\mu = a$ will
include $x=0$ and no null results are possible.
If $I$ is bigger than $\alpha$, the acceptance domain at $\mu = a$ will
NOT include 0 and null results will occur.

One thus sees that the occurence or not of null results with Feldman and
Cousins method will depend on the chosen CL. Contrary to the bayesian
approach, the absence of null results is not insured by construction.
\subsection{Some examples}
\subsubsection{exponential law}

$g(y) = e^{-y}$

$h(y)= y e^{-y}$  has a unique maximum at $y=1$.

$ I = \int_o^1 g(y) dy = 1 - e^{-1} = 0.63$

A CL higher than 0.63 will avoid null results.

This law is a $\chi_2^2$ law (2 degrees of freedom) for $2 y$.
More generally, $\chi_N^2$ laws give values of $I$ 
slowly decreasing to $1-3e^{-2}=0.594$ when
$N\rightarrow\infty$.

Other usual laws tend to give values of $I$ smaller than the usual choices of
CL, $0.9$ or higher, so that in practical cases, null results are likely to
be avoided. It is however possible, by a careful choice of $g(y)$, to obtain
situations where $I$ can be made as close to 1 as one wishes.
For example, the sigmoid-like function:
\[g(y) =\frac{k}{e^{(y-b)} + 1}\] with $y$ and $b$ positive 
gives for large values
of $b$ an integral $I$ whose value 
approaches $1 - \log(b)/b$ so that it can be
made higher than any given CL.

The next example, purely academic, exhibits a case where null results are 
unavoidable.
\subsubsection{flat distribution}

$g(y) = 1 $ for y between 0 and 1

$x$ in this case is flat between 0 and an unknown value $\mu$ restricted to
be bigger than $a$.
$r(x)$ in this case is equal to $a/\mu$ for $x\leq a$, and equal to $x/\mu$
for $x$ between $a$ and $\mu$. The acceptance domain for $x$ goes from
$(1-\alpha)\mu$ to $\mu$ (by connexity of the interval, higher values of
$x$ being selected first).

Thus, whatever the chosen value for $\alpha$, $x=0$ will be excluded from the
acceptance domain for all values of $\mu$, only lower limits
will be obtained and null results will always occur. 

\section{Discussion}
We have shown in this paper that the newly proposed method to put confidence
belts on bounded parameters fails to meet some of its advocated properties,
at least in principle although not in practice for most cases.

Why should such a method be preferred to others? (Note that PDG, in its
1998 edition \cite{PDG1998}, recommends this new method).
Feldman and Cousins argue that their method disantangles
estimation and hypothesis testing, but one can argue as well that they
obtain estimations from an 
ordering                                     
based on a quantity used for hypothesis testing!
In view of the failure to avoid null results,
I don't think any real
argument can be given if one wishes to stick to classical statistics.
In my opinion, different methods, as long as they give actual confidence
levels equal to the announced one, are mathematically equally acceptable, 
and a comparison between them implies, consciously or not, some
dose of "bayesian" input. 

The main problem at hand was to avoid null results. To cure this  
problem (completely 
for the bayesian solution, partly only with Feldman and Cousins), one is
led to break the symetry of central intervals which equally separates
the wrong statements between the victims of fluctuations towards high values of
$x$ and victims of fluctuations towards low values. To avoid null results,
a dissymetry is introduced to favor the latter to the detriment
of the former, who anyway will never know if they are right or wrong.
One could argue it is not fair!

I would like to add that the good points of Feldman and Cousins' method
could be translated to the bayesian approach to get rid of all the
criticisms it has received. We have already shown in section 2.4
how to solve the
criticisms of arbitrariness by using a modified probability density 
for $\mu$. The only critics left is that of overcoverage for bayesian
upper limits. But what Feldman and Cousins 
have shown is that one should publish intervals, which
happen to be upper limits when the interval starts at the physical
boundary. Thus, it would be perfectly legitimate, in order to avoid
overcoverage, to build the "$(1+\alpha)/2$" upper limit with our
modified bayesian recipee; it will show overcoverage, but this can be
removed by implementing a lower limit on $\mu$ obtained by defining
$x_{max}(\mu)$  so that the probability content of selected $x$ be $\alpha$
for any $\mu$. Such a construction would be devoid of any criticism\,:
no overcoverage, no null results, no arbitrariness, and it coïncides with the
classical central intervals of the Neyman construction for unbounded 
parameters.  
This is illustrated on figure~3.
\begin{figure}
\begin{center}
\mbox{\epsfig{file=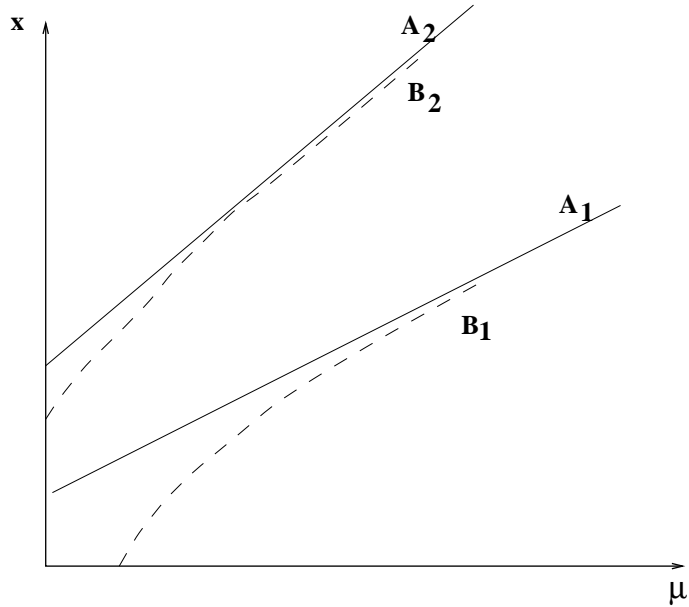,height=8cm}}
\end{center}
\caption{The classical belt at $\alpha$ CL is given by curves A1 and A2
defined by eqution~3. B1 is the bayesian upper limit obtained from formula~8
for a CL equal to $(1 + \alpha)/2$. It is complemented by the curve B2 so
that for each $\mu$ value, the probability content for $x$ between B1 and
B2 is $\alpha$.}
\end{figure}

The results obtained for Poisson statistics (in the case of a signal over
a known background) would be the same as the
usual $(1 + \alpha)/2$ (instead of $\alpha$, but this is the price
to be paid to avoid flip-flop biasing) bayesian upper limit 
suitably complemented by
a lower limit at high values of $x$, both curves 
coïnciding with the Neyman usual central interval for zero background.

This method I think would certainly satisfy the advocates of the bayesian
approach, and be acceptable by classical statisticians. Its construction
might look kind of odd, but it uses as well as others the freedom left
in the Neyman construction while fully responding to the anxiety created
(with no good reason in my opinion) by the occurence of null results.

I will close this paper with a final remark.

All the recent discussions on the choice of method to be used, linked to
the fact that different conclusions could be reached (for example, whether
or not the recent Karmen result~\cite{karmen} excludes the LSND
result~\cite{LSND} on neutrino oscillations \cite{Keitel}) 
have taken such an importance
due to the tendancy to publish results at lower and lower confidence levels.
Let me remind that the probability that 2 results being published at 90\% CL
are both right is only 0.81\,! And with 90 \% CL, a non negligible fraction 
of experiments can obtain null results. It would be in my opinion much
more reasonable to publish results with CL of at least 99\%, so that the
intensity of discussions would considerably drop, because on the one hand, 
the fraction of null results would reach a very low level and on the other
hand the 
probability of two results being both right would reach 98 \%, whatever the
algorithms used. Certainly results would look less spectacular, but
would gain in fiability.

\end{document}